\documentclass[prb,reprint,superscriptaddress]{revtex4-1}
\usepackage{amsmath,amssymb,bm,mathrsfs,graphicx, braket,amsthm,times}
\usepackage[colorlinks=true,citecolor=blue,linkcolor=blue]{hyperref}

\begin{document}

\title{Insensitivity of bulk properties to the twisted boundary condition} 

\author{Haruki Watanabe} 
\email{haruki.watanabe@ap.t.u-tokyo.ac.jp}
\affiliation{Department of Applied Physics, University of Tokyo, Tokyo 113-8656, Japan.}

\begin{abstract}
The symmetry and the locality are the two major sources of various general theorems in quantum many-body systems.  We demonstrate that, in gapped phases of a U(1) symmetric Hamiltonian with finite-range interactions, the bulk properties such as the expectation value of local operators, the ground state energy and the excitation gap, and the static and low-frequency dynamical responses in general, do not depend on the U(1) phase of the twisted boundary condition in the limit of the large system size. Specifically, their dependence on the twisted angle is exponentially suppressed with the linear dimension of the system.  Our argument is solely based on the exponential decay of various types of equal-time correlation functions and does not assume any details of the Hamiltonian, meaning that the statement applies quite generally regardless of the dimensionality or the interaction strength of the system.
\end{abstract}

\maketitle

\section{Introduction}
One of the main goals of theoretical condensed-matter physics is to achieve a systematic understanding of the interplay between symmetry and topology in many-body systems.  The topological properties of noninteracting band insulators can be characterized by various kinds of winding numbers, such as Berry phases and Chern numbers, of Bloch wave-functions as a function of the single-particle momentum~\cite{RMP_TI,RevModPhys.83.1057}.  This picture remains valid even when interactions are perturbatively taken into account~\cite{hohenadler2013correlation,XLQ2010,PhysRevB.83.085426,PhysRevB.87.121113}.  However, in the nonperturbative regime one needs an alternative approach. 

One possible solution to this problem can be formulated in terms of the twisted boundary condition.  This is a generalization of the more standard periodic boundary condition in which pairs of two surfaces in the opposite sides of the system are identified. In the twisted boundary condition, a U(1) phase is multiplied to one surface before being identified with its pair (the more precise definition is given in Sec.~\ref{sec:tbc}). The twisted phase $e^{i\theta_i}$ ($i=1,2,\ldots,d$) can be assigned independently for each direction.   It has been empirically known that the set of angles $\vec{\theta}=(\theta_1,\theta_2,\ldots,\theta_d)$ often serves as the many-body generalization of the single-particle momentum $\vec{k}=(k_1, k_2,\ldots,k_d)$. For example, the pumped charge in the Thouless pump~\cite{Thouless,NiuThouless} and the quantized Hall conductance of the quantum Hall effect~\cite{NTW,AvronSeiler} in interacting systems can be characterized by a Chern number formulated in terms of $\theta_i$ in stead of $k_i$.  There are also many studies defining the $\mathbb{Z}_2$ index for the many-body quantum spin Hall insulator using the twisted boundary condition~\cite{KaneMele,XLQ2006,Sheng2006,FuKane,RyuPRL,Hirano,XLQ2010,XGW2014,Chiu2016}. 

There is, however, a fundamental difference between $\vec{k}$ and $\vec{\theta}$.
The single-particle momentum $\vec{k}$ can be varied over the first Brillouin zone even under a fixed boundary condition.
Thus topological invariants written in terms of Bloch wavefunctions are properly defined for each Hamiltonian.  In contrast, varying $\theta$ changes the Hamiltonian itself, implying that the many-body topological invariants that involve integration(s) by $\theta_i$ are only defined for a series of Hamiltonians parametrized by $\vec{\theta}$.  For example, the Hall conductance $\sigma_{12}(\vec{\theta})$ can be computed using the linear response theory for each $\vec{\theta}=(\theta_1,\theta_2)$.  The quantization of this quantity nor its connection to Chern number is not obvious in this form.  The prescription proposed by Refs.~\onlinecite{NiuThouless,NTW} is to take an average of $\sigma_{12}(\vec{\theta})$ over all possible values of $\vec{\theta}$, \emph{assuming} that the $\vec{\theta}$-dependence of $\sigma_{12}(\vec{\theta})$ is negligibly small~\footnote{Ref.~\onlinecite{NTW} proposed a wrong power-law scaling ($\partial_{\theta_1}\sigma_{12}(\vec{\theta})\sim\frac{\xi}{L_1}$) in appendix, rather than the exponential suppression we show in this work.}. Then the resulting integral takes the form of the Chern number and the quantization to integers becomes apparent.   There are several recent studies that present an alternative proof of the quantization without performing such an average~\cite{HastingsMichalakis,Koma,Bachmann}. Note that the $\theta_i$-independence has been a common assumption behind countless subsequent works~\cite{AvronYaffe,Titus2011,LiuZhao,Zeng2015,Matsugatani}.

Another prominent application of the twisted boundary condition in the context of the topology in many-body systems is the generalization of the Lieb-Shultz-Mattis theorem~\cite{Lieb1961,Affleck1986,Affleck1988,OYA1997,Yamanaka} to multi-dimensions~\cite{Oshikawa2000,Hastings2004,NachtergaeleSims}.  The theorem states that, in a translation invariant system with the particle-number conservation, the filling (the average number of particles per unit cell) has to be an integer in order to realize a unique ground state with a nonzero excitation gap. An immediate consequence of this theorem is that any symmetric gapped phase with a fractional filling has to develop a ``topological order," which is usually accompanied by a fractionalization of particle statistics.  The first proof of the Lieb-Schultz-Mattis theorem in dimensions greater than one is given by Oshikawa~\cite{Oshikawa2000}, who interpreted the twist operator in the original one-dimensional argument~\cite{Lieb1961} as the large gauge translation operator. In the proof, he considered an adiabatic change of the twisted angle $\theta_x$ from $0$ to $2\pi$, \emph{assuming} that the excitation gap does not close in the process.  The recent refinement of the Lieb-Schultz-Mattis theorem in nonsymmorphic space groups~\cite{Sid2013,PNAS,WatanabePRB} essentially relies on the same assumption.  In fact, the stability of the gap against an increase of $\theta_x$ is, in general, not at all for granted. For example, the excitation gap in the Kitaev chain vanishes at some values of $\theta_x$~\cite{Kawabata}.  Hastings then gave an alternative proof without such an assumption~\cite{Hastings2004,NachtergaeleSims}, but instead relying on a `reality condition'~\footnote{See \emph{Condition LSM6} in Sec. 1.2 of Ref.~\onlinecite{NachtergaeleSims} that corresponds to the footnote [19] in Ref~\onlinecite{Hastings2004}.}.

To summarize, the $\theta$-independence of the bulk properties such as the excitation gap and the linear response coefficient have been an assumption in pioneering studies on the many-body topological invariants and the multi-dimensional Lieb-Schultz-Mattis theorem.  Although there have been follow-up works that discuss an alternative derivation that goes around the assumption for each problem, it would be nicer to have a general and direct verification of the assumption itself, as it may lead to new applications of the twisted boundary condition.  In this paper we give a general proof of the  insensitivity of bulk properties to the twisted angle $\vec{\theta}$, assuming (i) the locality and the U(1) symmetry of the Hamiltonian and (ii) a non-zero excitation gap and the uniqueness of the ground state for one value of $\vec{\theta}$ (e.g, $\vec{\theta}=\vec{0}$). Our argument coherently applies to expectation values [see Eq.~\eqref{eq:statement1}], static susceptibilities, the Thouless pump and the Hall conductance, and many other bulk response properties [see Eq.~\eqref{eq:statement2}]. 
As a by-product, we prove the exponential decay of several new types of correlation functions [see Eqs.~\eqref{decay2}, \eqref{bounddecay3}, and  \eqref{bounddecay4}].

The organization of this paper is as follows. In Sec.~\ref{sec:decay}, we summarize the general behavior of correlation functions in gapped phases. In Sec.~\ref{sec:tbc}, we review the definition of the twisted boundary condition and its understanding in terms of the magnetic flux.  
With these preparations, we prove that various quantities in many-body systems do not depend on the twisted angle of the boundary condition in the limit of a large system size. We start from the expectation value of charge-conserving operators in Sec.~\ref{sec:exp}, and then move on to the static responses and topological transport properties in Sec.~\ref{sec:response}, and finally discuss the excitation gap in Sec.~\ref{sec:gap}.
Then we conclude in Sec.~\ref{sec:conclude}.

\section{Exponential decay of correlation functions}
\label{sec:decay}

\subsection{Assumptions: the locality and the gap}
Consider a quantum system in $d$ spatial dimensions.  To discuss a finite-size system without a boundary, we impose the periodic boundary condition with the linear dimension $L_i$ in $i$-th direction ($i=1,2,\ldots,d$).  Suppose that the Hamiltonian $\hat{H}$ of the system is given as a sum of \emph{local} terms: 
\begin{equation}
\label{eq:hamiltonian}
\hat{H}=\sum_{\vec{x}}\,\hat{H}_{\vec{x}}.
\end{equation}
We say $\hat{H}_{\vec{x}}$ is local when its range $r$ is finite and does not scale with the system size.  Namely, $\hat{H}_{\vec{x}}$ does not affect the local Hilbert space at $\vec{y}$ whenever $|\vec{y}-\vec{x}|>r$~\footnote{The assumption of finite-range interactions could be relaxed to exponential decaying or even to algebraic decaying interactions. This, in turn, requires more careful and mathematically elaborated treatment~\cite{HastingsKoma}.}.  For example, the term $\hat{H}_{\vec{x}}=\sum_{\vec{y}}t_{\vec{x},\vec{y}}c_{\vec{x}}^\dagger c_{\vec{y}}+\text{h.c.}$ in the tight-binding model is local when $t_{\vec{x},\vec{y}}=0$ for $|\vec{y}-\vec{x}|>r$.  The support of an operator $\hat{H}_{\vec{x}}$ is the set of $\vec{y}$ at which $\hat{H}_{\vec{x}}$ acts nontrivially. Thus, the support is a subset of the ``ball" with the radius $r$ centered at $\vec{x}$.  For continuum model, the sum in Eq.~\eqref{eq:hamiltonian} should be replaced by an integral.

Throughout the paper, we assume that the ground state $|0\rangle$ of $\hat{H}$ is unique and that the excitation gap $\Delta$ does not vanish in the limit of large system size.  We will comment on the case with a finite ground-state degeneracy at the end of the paper.  We focus on zero temperature $T=0$ and $\langle\hat{O}\rangle$ denotes the expectation value $\langle 0|\hat{O}|0\rangle$ with respect to the ground state. Furthermore, $\delta\hat{O}$ represents the fluctuation $\hat{O}-\langle \hat{O}\rangle$ and the time-evolution of an operator is defined by $\hat{O}(t)\equiv e^{i\hat{H}t}\hat{O}e^{-i\hat{H}t}$. 

\subsection{The behavior of correlation functions}
\begin{figure}
\begin{center}
\includegraphics[width=0.99\columnwidth]{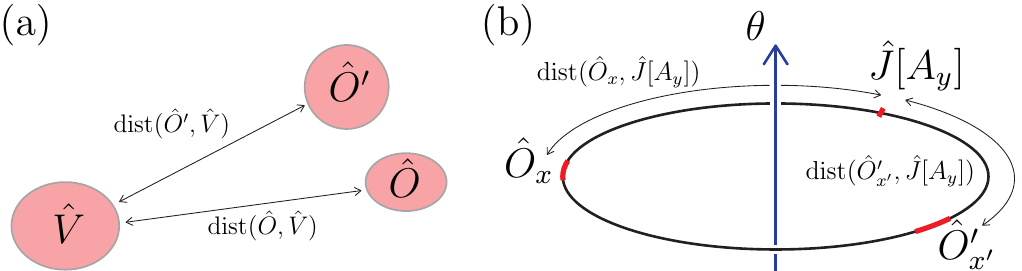}
\caption{\label{fig} (a) The spatial configuration of local operators $\hat{O}$, $\hat{O}'$ and  $\hat{V}$. The red shades represent their support.  (b) The flux $\theta$ piercing the ring. }
\end{center}
\end{figure}

Let $\hat{O}$ and $\hat{V}$ be local operators and let $R\equiv\text{dist}(\hat{O},\hat{V})$ be the minimum distance between their support [Fig.~\ref{fig} (a)]. We assume $R>0$; in other words, the support of $\hat{O}$ and $\hat{V}$ do not overlap.  In gapped phases, it is well known, and is also rigorously proven~\cite{HastingsPRL,HastingsKoma}, that the equal-time (connected) correlation function decays exponentially with the distance:
\begin{equation}
F_0\equiv\langle\delta\hat{O}\,\delta\hat{V}\rangle,\quad |F_0| \leq C_0e^{-\frac{R}{\xi}}.\label{decay1}
\end{equation}
In fact, a similar argument proves that the correlation function of the following form also decays exponentially
\begin{equation}
F_{n}\equiv\langle\delta\hat{O}\frac{1}{(\hat{H}-E)^{n}}\delta\hat{V}\rangle,\quad  |F_{n}|\leq C_n R^{\frac{n}{2}}e^{-\frac{R}{\xi}}\label{decay2}.
\end{equation}
Here, $n=1,2,\ldots$ is an arbitrary natural number and $E$ is the ground state energy. The proof for $F_{2}$ can be found in Ref.~\onlinecite{Koma}, although it is buried in a long mathematically-elaborated paper.  In Appendix A, we present the simplest version of the proof in a way applicable to all $n$.  The key tool of the proof is the Lieb-Robinson bound~\cite{LRB,HastingsReview}
\begin{equation}
\label{eq:LRB}
\|[\hat{O},\hat{V}(t)]\|\leq Ce^{-\frac{R}{\xi_0}}(e^{\frac{v|t|}{\xi_0}}-1).
\end{equation}
Here $\|\hat{O}\|\equiv\text{sup}_{|\psi\rangle, \langle\psi|\psi\rangle=1}\|\hat{O}|\psi\rangle\|$ denotes the norm of the operator $\hat{O}$, and the constants $\xi_0$ and $v$ are dependent on the Hamiltonian $\hat{H}$ but are independent of the choice of operators $\hat{O}$ or $\hat{V}$.  The Lieb-Robinson bound intuitively estimates the spreads of the operator $\hat{V}(t)$ as the time evolves. For example, at $t=0$, the right-hand side of Eq.~\eqref{eq:LRB} vanishes. This is because operators $\hat{O}$ and $\hat{V}$ themselves commute (recall our assumption of $R>0$). As the time grows, the support of the operator $\hat{V}(t)$ expands and overlaps with the support of $\hat{O}$. The Lieb-Robinson bound gives the upper limit of the velocity $v$ of this spread. 

The correlation length $\xi$ in Eqs.~\eqref{decay1} and \eqref{decay2} is given by $\xi\equiv\xi_0+\frac{2v}{\Delta}$, where the constants $\xi_0$ and $v$ are those appearing in Eq.~\eqref{eq:LRB}.  When the gap $\Delta$ is small, the correlation length $\xi$ is dominated by $\frac{2v}{\Delta}$ and diverges in the limit of $\Delta\rightarrow+0$ as expected.

The correlation functions in Eqs.~\eqref{decay1} and \eqref{decay2} are about two operators at a distance. Let us now consider correlations involving more operators, e.g., $G_{00}\equiv\langle\delta\hat{O}\,\delta\hat{O}'\,\delta\hat{V}\rangle$.  We assume that the support of $\hat{V}$ is well separated from that of $\hat{O}$ and $\hat{O}'$, while assuming nothing about the distance between the support of $\hat{O}$ and $\hat{O}'$ [Fig.~\ref{fig} (a)]. In this case, one can simply regard the product $\hat{O}\hat{O}'$ as a single operator and apply  Eq.~\eqref{decay1} to get a bound $|G_{00}|\leq C_{00}e^{-\frac{R'}{\xi}}$, where $R'$ is either the smaller one of $\text{dist}(\hat{O},\hat{V})$ and $\text{dist}(\hat{O}',\hat{V})$. In contrast, the following correlations cannot be evaluated directly through Eqs.~\eqref{decay1} or \eqref{decay2},
\begin{eqnarray}
G_{mn}&\equiv&\langle\delta\hat{O}\frac{1}{(\hat{H}-E)^{m}}\delta\hat{O}'\frac{1}{(\hat{H}-E)^{n}}\delta\hat{V}\rangle,\label{decay3}\\
G'_{mn}&\equiv&\langle\delta\hat{O}\frac{1}{(\hat{H}-E)^{m}}\delta\hat{V}\frac{1}{(\hat{H}-E)^{n}}\delta\hat{O}'\rangle,\label{decay4}
\end{eqnarray}
because the product $\hat{O}\frac{1}{(\hat{H}-E)^{m}}\hat{O}'$ ($m=1,2,\ldots$) is not necessarily local even when $\hat{O}$ and $\hat{O}'$ are~\footnote{This is because $(\hat{H}-E)^{-m}$ is not necessarily a sum of local operators. This might be understood easily by recalling that, even if a matrix $M$ is almost diagonal, $M^{-1}$ can be non-diagonal at all.}.  Nevertheless, we can prove (see Appendix B)
\begin{eqnarray}
|G_{mn}|&\leq& C_{mn} {R'}^{\frac{n+m}{2}}e^{-\frac{R'}{\xi'}},\label{bounddecay3}\\
|G_{mn}'|&\leq& C_{mn}' {R'}^{\frac{n+m+1}{2}}e^{-\frac{R'}{\xi'}},\label{bounddecay4}
\end{eqnarray}
where $\xi'\equiv\xi_0+\frac{4v}{\Delta}$ and $R'$ is defined above Eq.~\eqref{decay3}.

\subsection{Perturbation at distance}
The properties of correlation functions summarized above have many valuable implications which do not seem fully explored.  As a simple example, let us show that any perturbation at a long distance never affects the expectation value of a local operator. We consider a Hamiltonian $\hat{H}(h)=\hat{H}-h\hat{V}$ with a local perturbation $\hat{V}$.  Let $|h\rangle$ be the unique ground state $\hat{H}(h)$.  Then, differentiating the defining equation $\hat{H}(h)|h\rangle=E(h)|h\rangle$, one gets
\begin{equation}
\hat{Q}(h)\partial_h|h\rangle=-\frac{1}{\hat{H}(h)-E(h)}\delta\hat{V}|h\rangle,\label{derivative}
\end{equation}
where $\hat{Q}(h)\equiv1-|h\rangle\langle h|$ is the projection onto excited states.  For the expectation value $O(h)\equiv\langle h|\hat{O}|h\rangle$ of a local Hermitian operator $\hat{O}$, the derivative $\partial_hO(h)$ is thus given in the form of $F_1$:
\begin{equation}
\partial_hO(h)=-\langle h|\delta\hat{O}\frac{1}{\hat{H}(h)-E(h)}\delta\hat{V}|h\rangle+\text{c.c.},\label{susceptibility}
\end{equation}
which is exponentially small when $\hat{O}$ and $\hat{V}$ are well-separated, as suggested by Eq.~\eqref{decay2}:
\begin{equation}
|\partial_hO(h)|\leq C \sqrt{R}e^{-\frac{R}{\xi}}
\end{equation}
where $C$ is a constant and $R$ is the distance between $\hat{O}$ and $\hat{V}$.

\section{Twisted boundary condition and U(1) symmetry}
\label{sec:tbc}
As a preparation for discussing more nontrivial applications of the exponential decay of correlation functions, in this section we review the basics of the twisted boundary condition and its connection to magnetic flux.

\subsection{Twisted boundary condition}
\label{subsec:tbc}
Suppose that the Hamiltonian $\hat{H}=\int d^dx\,\hat{H}_{\vec{x}}$ is written in terms of the creation (annihilation) operator $\hat{c}_{\vec{x}}^\dagger$ ($\hat{c}_{\vec{x}}$). The total number operator $\hat{N}\equiv\int d^dx\,\hat{n}_{\vec{x}}$ is the integral of the number density operator $\hat{n}_{\vec{x}}\equiv\hat{c}_{\vec{x}}^\dagger \hat{c}_{\vec{x}}$ and the global U(1) phase rotation is described by $e^{i\phi\hat{N}}$.

Let $\hat{T}_{\vec{v}}$ be the operator that describes the translation by $\vec{v}$ and let $\hat{x}_i$ be the unit vector along the $i$-th axis ($i=1,2,\ldots,d$) of the Cartesian coordinate. Recall that the periodic boundary condition is set by identifying two surfaces $x_i=0$ and $x_i=L_i$. In other words, we identify the translation operator $\hat{T}_{L_i\hat{x}_i}$ as the identity operator:
\begin{equation}
\hat{T}_{L_i\hat{x}_i}=1.
\end{equation}
The extension to the twisted boundary condition can be done simply by setting instead the product of the translation operator $\hat{T}_{L_i\hat{x}_i}$ and the phase rotation operator $e^{i\theta_i\hat{N}}$ as the identity:
\begin{equation}
\label{eq:tbc}
\hat{T}_{L_i\hat{x}_i}e^{i\theta_i\hat{N}}=1.
\end{equation}
Under this identification, the creation operator $\hat{c}_{\vec{x}}^\dagger$, for example, satisfies 
\begin{equation}
\hat{c}_{\vec{x}+L_i\hat{x}_i}^\dagger=e^{-i\theta_i}\hat{c}_{\vec{x}}^\dagger
\end{equation}
for every $i=1,2,\ldots,d$. We denote by $\hat{H}(\vec{\theta})$ the resulting Hamiltonian written in terms of operators $\hat{c}_{\vec{x}}^\dagger$ and $\hat{c}_{\vec{x}}$ in the range $\vec{x}\in[0,L_1)\times[0,L_2)\times\ldots\times[0,L_d)$.

\subsection{U(1) symmetry and magnetic flux}
\label{subsec:u1}
There is a distinct but equivalent view of $\theta_i$ in terms of the magnetic flux when the Hamiltonian has the global U(1) symmetry.   Let us start with the Hamiltonian under the periodic  boundary condition $\hat{H}(\vec{0})$. 
Let us consider a unitary operator $\hat{U}_\chi\equiv e^{i\int d^dx\,\chi(\vec{x})\hat{n}_{\vec{x}}}$ that multiplies a position-dependent phase $e^{i\chi(\vec{x})}$ to $\hat{c}_{\vec{x}}^\dagger$. Here, $\chi(\vec{x})$ is an arbitrary piecewise smooth function of $\vec{x}$, and the Hamiltonian is not necessarily invariant under such a local U(1) rotation.  We introduce a \emph{non-dynamical} gauge field $\vec{A}(\vec{x})$ in such a way that (i) $\hat{H}[\vec{A}]=\int d^dx\,\hat{H}_{\vec{x}}[\vec{A}]$ transform as
\begin{equation}
\hat{U}_\chi\hat{H}_{\vec{x}}[\vec{A}]\hat{U}_\chi^\dagger=\hat{H}_{\vec{x}}[\vec{A}'],\quad \vec{A}'(\vec{x})\equiv \vec{A}(\vec{x})-\partial_{\vec{x}}\chi(\vec{x})\label{Htrans}
\end{equation}
and (ii) $\hat{H}[\vec{A}]$ reduce to $\hat{H}(\vec{0})$ when $\vec{A}(\vec{x})=\vec{0}$.  We can always introduce $\vec{A}$ with this property as long as the Hamiltonian $\hat{H}(\vec{0})$ has the global U(1) symmetry (i.e., commutes with the number operator $\hat{N}$).  The simplest example of $\hat{H}_{\vec{x}}[\vec{A}]$ may be 
\begin{equation}
\hat{H}_{\vec{x}}[A]=\hat{c}_{\vec{x}}^\dagger\left[-\tfrac{1}{2m}(\partial_{\vec{x}}+i\vec{A}(\vec{x}))^2+U(\vec{x})\right]\hat{c}_{\vec{x}}+\hat{H}_{\vec{x}}^{\text{int}},
\end{equation}
where $U(x)$ is the single particle potential and $\hat{H}_{x}^{\text{int}}$ describes the many-body interactions.  A bad example would be the (meanfield) BCS Hamiltonian which lacks the U(1) symmetry due to the presence of terms proportional to $\hat{c}\hat{c}$ or $\hat{c}^\dagger \hat{c}^\dagger$.  In this case, there is no way to introduce $\vec{A}(x)$ satisfying Eq.~\eqref{Htrans}.  

We describe the ``magnetic flux" $\theta_i\equiv\int_0^{L_i} dx_i A_i(x)$ by choosing a position-independent vector potential
\begin{equation}
\vec{A}(\vec{x})=(\tfrac{\theta_1}{L_1},\tfrac{\theta_2}{L_2},\ldots,\tfrac{\theta_d}{L_d}).
\end{equation}
We write the resulting Hamiltonian as $\hat{H}'(\vec{\theta})=\hat{H}[\vec{A}]$. Note that we did not actually apply any real ``magnetic field" to the system.  The magnetic flux $\theta_i$ is pierced through the hole of the ``ring" formed by the $x_i$ axis under the boundary condition identifying $x_i=L_i$ and $x_i=0$. See Fig.~\ref{fig} (b) for the illustration in the case of $d=1$.

\subsection{Equivalence of $\hat{H}(\vec{\theta})$ and $\hat{H}'(\vec{\theta})$}
The Hamiltonian $\hat{H}(\vec{\theta})$ under the twisted boundary condition in Sec.~\ref{subsec:tbc} and the Hamiltonian $\hat{H}'(\vec{\theta})$ under the magnetic flux in Sec.~\ref{subsec:u1} are, in fact, unitary equivalent with each other.  Therefore, they describe physically the same system; in particular, their spectrum and the properties of correlation functions, for example, are the same.  The two Hamiltonians are related by $\hat{U}_{\chi}$ with  $\chi(\vec{x})=\sum_{i=1}^d\theta_i\frac{x_i}{L_i}$:
\begin{equation}
\hat{U}_{\chi}\hat{H}'(\vec{\theta})\hat{U}_{\chi}^\dagger=\hat{H}(\vec{\theta}).
\end{equation}
Note that the function $\chi(\vec{x})$ is discontinuous at the boundary jumping from $\theta_i$ at $x_i=L_i$ to $0$ at $x_i=0$.  Using Eq.~\eqref{Htrans}, we find that
\begin{equation}
A_i'(\vec{x})=\tfrac{\theta_i}{L_i}-\partial_{x_i}\chi(\vec{x})=\theta_i\delta(x_i),
\end{equation}
where the $\delta$-function originates from the discontinuity of $\chi$ at the boundary.  This means that the Hamiltonian $\hat{H}(\vec{\theta})$ under the twisted boundary condition can be interpreted as the Hamiltonian subjected to the $\delta$-function-type vector potential localized at the boundary.  This should also clarifies that we can freely move the position of the $\delta$-function peak in the system by performing a proper gauge transformation. This s is actually what we do in the following sections [e.g., see Eq.~\eqref{gaugey}].

\section{Insensitivity of expectation values}
\label{sec:exp}
With these preparations, let us now demonstrate that the expectation value of a wide class of operators do not depend on $\vec{\theta}$ in the limit of large $L_i$.  
To simplify the notation here we focus on 1D systems (and thus drop the subscript ``1"). This is actually sufficient to prove the same claim in higher dimensions since we can apply the 1D argument for each direction separately.

Let us consider an operator $\hat{O}=\int_0^{L}dx\,\hat{O}_x$ that is given as an integral of local terms $\hat{O}_x$ and commutes with $\hat{N}$.
We can then introduce $A$ so that $\hat{O}[A]=\int_0^{L}dx\,\hat{O}_x[A]$ transforms in the same way as $\hat{H}[A]$ does in Eq.~\eqref{Htrans}.  The operator $\hat{O}$ can be the Hamiltonian $\hat{H}$ itself, but it may also be, for example, the polarization operator $\hat{P}=\int_0^{L}dx\,x\hat{n}_x$ or the current operator.

Now we choose the uniform vector potential $A(x)=\frac{\theta}{L}$.  We denote the unique ground state of $\hat{H}'(\theta)=\hat{H}[\frac{\theta}{L}]$ by $|\theta\rangle$.  Our claim is that the $\theta$-dependence of the expectation value 
\begin{equation}
O(\theta)\equiv\langle\theta|\hat{O}[\tfrac{\theta}{L}]|\theta\rangle=\int_0^{L} dx\langle\theta|\hat{O}_x[\tfrac{\theta}{L}]|\theta\rangle\label{expectation}
\end{equation}
is suppressed for a large $L$ by a factor $L^{3/2}e^{-\frac{L}{2\xi}}$.   When $\hat{O}=\hat{H}$, the statement is the flatness of the ground state energy $E_\theta$ as a function of $\theta$, which was numerically observed before, e.g., in Ref.~\onlinecite{Misguich}.  Later we will also argue that the excitation gap is independent of $\theta$ in the limit of large $L$.

To prove the claim, let us define a function of $x$ labeled by $y\in[0,L]$. It reads 
\begin{equation}
\label{gaugey}
\chi_{y}(x)=
\begin{cases}
\tfrac{\theta}{L}x& (0\leq x<y)\\
\tfrac{\theta}{L}(x-L)& (y\leq x<L).
\end{cases}
\end{equation}
The corresponding unitary operator $\hat{U}_{\chi_{y}}=e^{i\int_0^{L}dx\,\chi_{y}(x)\hat{n}_x}$ induces the gauge transformation
\begin{equation}
A(x)=\tfrac{\theta}{L}\,\,\,\rightarrow\,\,\,A_{y}(x)\equiv \theta\delta(x-y).\label{Gauge}
\end{equation}
In this gauge, one can say $\theta$ is the U(1) phase of the twisted boundary condition at the new boundary $x=y$.

The key observation is that, thanks to the assumed locality, \emph{$\hat{O}_x[A_y]$ is independent of $\theta$} and thus is identical to $\hat{O}_x[0]$ when $y$ is out of the range of $\hat{O}_x$.  Namely, if we denote by $r$ the maximum range of $\hat{O}_x$ over all $x\in[0,L]$, then we have
\begin{equation}
\hat{O}_x[A_y]=\hat{O}_x[0]\quad\text{if}\quad|y-x|>r.\label{eq:trick}
\end{equation}  
For example, in the case of $\hat{O}_x[A]=t\hat{c}_{x+r}^\dagger e^{-i\int_{x}^{x+r}dz A(z)}\hat{c}_x$, 
\begin{equation}
\hat{O}_x[A_y]=t\hat{c}_{x+r}^\dagger e^{-i\theta\int_{x}^{x+r}dz \delta(z-y)}\hat{c}_x=t\hat{c}_{x+r}^\dagger\hat{c}_x
\end{equation}
is independent of $\theta$ as long as $|y-x|>r$.  It follows that the local terms of the Hamiltonian $\hat{H}_x[A_{y}]$ do not depend on $\theta$ either unless $y$ is within the range of $\hat{H}_x$.

Inserting $\hat{U}_{\chi_{y}}^\dagger\hat{U}_{\chi_{y}}=1$ to the last expression in Eq.~\eqref{expectation} and writing $|\theta_{y}\rangle\equiv\hat{U}_{\chi_{y}}|\theta\rangle$, we get
\begin{equation}
O(\theta)=\int_0^{L}dx\,\langle\theta_{y}|\hat{O}_x[A_y]|\theta_{y}\rangle.
\end{equation}
Note that the value of $y$ here is arbitrary and can be chosen \emph{depending on $x$}. Thus we can freely set $y$ to be far away from $x$ so that $\hat{O}_x[A_y]=\hat{O}_x[0]$  [Fig.~\ref{fig} (b)]. For example, take the opposite point of $x$ on the ring with $|x-y|=\frac{L}{2}$:
\begin{equation}
O(\theta)=\int_0^{L}dx\,\langle\theta_{y}|\hat{O}_x[0]|\theta_{y}\rangle,\quad |x-y|=\tfrac{L}{2}>r.
\end{equation}
Then, intuitively, the twisted angle $\theta$ does not affect the expectation value $\langle\theta_{y}|\hat{O}_x[0]|\theta_{y}\rangle$ since $|\theta_{y}\rangle=\hat{U}_{\chi_{y}}|\theta\rangle$ is the ground state of $\hat{H}[A_{y}]$ twisted only near $y$, far away from $x$.
In fact, using Eq.~\eqref{derivative} for $h=\theta$, we can express $\partial_\theta O(\theta)$ in the form of $F_1$:
\begin{eqnarray}
\label{eq:step}
\partial_\theta O(\theta)=&&-\int_0^{L} dx\Big(\langle\theta_{y}|\delta\hat{O}_x[0]\frac{1}{\hat{H}[A_{y}]-E_\theta}\delta\hat{J}[A_{y}]|\theta_{y}\rangle\notag\\
&&\quad\quad+\langle\theta_{y}|\delta\hat{J}[A_{y}]\frac{1}{\hat{H}[A_{y}]-E_\theta}\delta\hat{O}_x[0]|\theta_{y}\rangle\Big),
\end{eqnarray}
where
\begin{equation}
\hat{J}[A_{y}]\equiv \partial_\theta\hat{H}[A_{y}]
\end{equation}
is the local current operator at $y$. Therefore, one can apply Eq.~\eqref{decay2} for $R=\frac{L}{2}$ to the integrand and get the desired bound
\begin{equation}
|\partial_\theta O(\theta)|< C L^{3/2}e^{-\frac{L}{2\xi}}
\end{equation}
with a constant $C$. In a higher dimension, the same argument leads to
\begin{equation}
\label{eq:statement1}
|\partial_{\theta_i} O(\vec{\theta})|< C V L_i^{1/2}e^{-\frac{L_i}{2\xi}}
\end{equation}
for each direction $i=1,2,\dots,d$. Here, $V=L_1\cdots L_d$ is the volume of the system, which originates from the integral in Eq.~\eqref{eq:step}.

\section{Insensitivity of bulk responses}
\label{sec:response}
Let us move on to the discussion of $\vec{\theta}$-independence of bulk responses. Specifically, we will focus on the class of responses that can be characterized by the correlation function of the form
\begin{equation}
G_n(\theta)=\langle\theta|\delta\hat{O}[\tfrac{\theta}{L}]\frac{1}{(\hat{H}[\tfrac{\theta}{L}]-E_\theta)^{n}}\delta\hat{O}'[\tfrac{\theta}{L}]|\theta\rangle.
\end{equation}
For example, the \emph{static} susceptibility, in general, takes the form $G_1(\theta)$ as demonstrated in Sec.~\ref{sec:decay} (see Eq.~\eqref{susceptibility}). The simplest instance is the static magnetic susceptibility corresponding to the choice $\hat{O}=\hat{O}'=\hat{S}_z$.  As we will see now, the correlation $G_2(\theta)$ is related to topological transports.  

\subsection{Thouless pump}
When the Hamiltonian has an adiabatic and periodic time dependence, the phenomenon so-called Thouless pump takes place and a certain amount of charge is transported through the system over time.  According to Ref.~\onlinecite{NiuThouless}, the pumped charge of a weakly time-dependent Hamiltonian over one cycle $T$ is given by 
\begin{equation}
\Delta Q(\theta)=i\int_{0}^Tdt (\partial_t\langle\theta|\partial_\theta|\theta\rangle-\partial_\theta\langle\theta|\partial_t|\theta\rangle).\label{ThoulessPump}
\end{equation}
Here $|\theta\rangle$ is the ground state of the snapshot Hamiltonian $\hat{H}[\frac{\theta}{L}]$.  Using Eq.~\eqref{derivative}, we can rewrite $\Delta Q(\theta)$ in the form of $G_2(\theta)$:
\begin{eqnarray}
\Delta Q(\theta)=&&i\int_{0}^Tdt\Big(\langle\theta|\delta(\partial_t\hat{H}[\tfrac{\theta}{L}])\frac{1}{(\hat{H}[\tfrac{\theta}{L}]-E_\theta)^{2}}\delta\hat{J}[\tfrac{\theta}{L}]|\theta\rangle\notag\\
&&\quad-\langle\theta|\delta\hat{J}[\tfrac{\theta}{L}]\frac{1}{(\hat{H}[\tfrac{\theta}{L}]-E_\theta)^{2}}\delta(\partial_t\hat{H}[\tfrac{\theta}{L}])|\theta\rangle
\Big).\label{ThoulessPump2}
\end{eqnarray}
Note that the $\theta$-integral is missing in Eqs.~\eqref{ThoulessPump} and \eqref{ThoulessPump2}.

\subsection{Hall conductance}
The Hall conductance can be formulated in a similar manner. Following Refs.~\onlinecite{NTW,AvronSeiler}, let us introduce the constant vector potential $\vec{A}(x,y)=(\frac{\theta_1}{L_1},\frac{\theta_2}{L_2})$.  If we denote by $|\vec{\theta}\rangle$ the ground state of $\hat{H}[\vec{A}]$, the Hall conductance is given by~\cite{NTW,AvronSeiler}
\begin{equation}
\sigma_{12}(\vec{\theta})=\frac{e^2}{h}2\pi i(\partial_{\theta_2}\langle\vec{\theta}|\partial_{\theta_1}|\vec{\theta}\rangle-\partial_{\theta_1}\langle\vec{\theta}|\partial_{\theta_2}|\vec{\theta}\rangle),\label{Hall}
\end{equation}
which can be written in the form of $G_2(\vec{\theta})$ using Eq.~\eqref{derivative}:
\begin{eqnarray}
\sigma_{12}(\vec{\theta})=&&\frac{e^2}{h}2\pi i\Big(\langle\vec{\theta}|\delta\hat{J}_2[\vec{A}]\frac{1}{(\hat{H}[\vec{A}]-E_{\vec{\theta}})^{2}}\delta\hat{J}_1[\vec{A}]|\vec{\theta}\rangle\notag\\
&&\quad-\langle\vec{\theta}|\delta\hat{J}_1[\vec{A}]\frac{1}{(\hat{H}[\vec{A}]-E_{\vec{\theta}})^{2}}\delta\hat{J}_2[\vec{A}]|\vec{\theta}\rangle
\Big).\label{Hall2}
\end{eqnarray}
Again, $\theta_{1,2}$ integrals are missing in Eqs.~\eqref{Hall} and \eqref{Hall2}, although they are the key in identifying this quantity as the Chern number. As we will show in Sec.~\ref{sec:proofGn}, $G_n(\theta)$ is almost independent of $\theta$ in a large system. Thus one can approximate $\sigma_{12}(\vec{\theta})$ by its average $\bar{\sigma}_{12}$~\cite{NTW}:
\begin{eqnarray}
\bar{\sigma}_{12}&\equiv&\int_0^{2\pi}\frac{d\theta_1}{2\pi} \int_0^{2\pi}\frac{d\theta_2}{2\pi} \sigma_{12}(\vec{\theta})=\frac{e^2}{h}C,\\
C&\equiv&\int\frac{d^2\theta}{2\pi}i(\partial_{\theta_2}\langle\vec{\theta}|\partial_{\theta_1}|\vec{\theta}\rangle-\partial_{\theta_1}\langle\vec{\theta}|\partial_{\theta_2}|\vec{\theta}\rangle).
\end{eqnarray}
The connection to the Chern number is now evident~\cite{NTW}.  We can perform the same trick to $\Delta Q(\theta)$ and relate it to a Chern number in the $t$-$\theta$ space~\cite{NiuThouless}.

\subsection{Insensitivity of $G_n(\theta)$}
\label{sec:proofGn}
Motivated by these examples, let us now prove that the $\theta$-dependence of $G_n(\theta)$ is exponentially suppressed for a large system by a factor $L^{2+\frac{n}{2}}e^{-\frac{L}{4\xi'}}$ with $\xi'\equiv\xi_0+\frac{4v}{\Delta}$. Our proof proceeds in the same way as that for $O(\theta)$.  Again we focus on one dimension. 

We first write $G_n(\theta)$ in terms of the integral of local operators 
\begin{equation}
G_n(\theta)=\int dxdx'\langle\theta|\delta\hat{O}_x[\tfrac{\theta}{L}]\frac{1}{(\hat{H}[\tfrac{\theta}{L}]-E_\theta)^{n}}\delta\hat{O}_{x'}'[\tfrac{\theta}{L}]|\theta\rangle
\end{equation}
and then insert $\hat{U}_{\chi_{y}}^\dagger\hat{U}_{\chi_{y}}=1$:
\begin{equation}
G_n(\theta)=\int dxdx'\langle\theta_y|\delta\hat{O}_x[0]\frac{1}{(\hat{H}[A_y]-E_\theta)^{n}}\delta\hat{O}_{x'}'[0]|\theta_y\rangle.\label{Gnp}
\end{equation}
In Eq.~\eqref{Gnp}, we have chosen $y\in[0,L]$ to be out of the range of $\hat{O}_x$, $\hat{O}_{x'}'$ as illustrated in Fig.~\ref{fig} (b) and used Eq.~\eqref{eq:trick}.  In fact, for every $x,x'\in [0,L]$, we can always find $y$ on the ring such that $|x-y|\geq \frac{L}{4}$ and $|x'-y|\geq \frac{L}{4}$.  Again using Eq.~\eqref{derivative}, we can express $\partial_\theta G_n(\theta)$ in terms of  $G_{m,\ell}$ and $G_{m,\ell}'$ with $m+\ell=n+1$:
\begin{widetext}
\begin{eqnarray}
\partial_\theta G_n(\theta)=-\int_0^{L}dx\int_0^{L}dx'&&\Big(\sum_{m=1}^{n}\langle\theta_y|\delta\hat{O}_x[0]\frac{1}{(\hat{H}[A_y]-E_\theta)^{m}}\delta\hat{J}[A_{y}]\frac{1}{(\hat{H}[A_y]-E_\theta)^{n-m+1}}\delta\hat{O}_{x'}'[0]|\theta_z\rangle\notag\\
&&\quad\quad\quad\quad+\langle\theta_y|\delta\hat{O}_x[0]\frac{1}{(\hat{H}[A_y]-E_\theta)^n}\delta\hat{O}_{x'}'[0]\frac{1}{\hat{H}[A_y]-E_\theta}\delta\hat{J}[A_{y}]|\theta_y\rangle\notag\\
&&\quad\quad\quad\quad+\langle\theta_y|\delta\hat{J}[A_z]\frac{1}{\hat{H}[A_y]-E_\theta}\delta\hat{O}_x[0]\frac{1}{(\hat{H}[A_y]-E_\theta)^n}\delta\hat{O}_{x'}'[0]|\theta_y\rangle\Big).
\end{eqnarray}
\end{widetext}
Thus one can use Eqs.~\eqref{bounddecay3} and \eqref{bounddecay4} with $R'=\frac{L}{4}$ to get the stated bound,
\begin{equation}
|\partial_\theta G_n(\theta)|< C L^{3+\frac{n}{2}}e^{-\frac{L}{4\xi'}}
\end{equation}
with a constant $C$. In a higher dimension, the same argument suggests
\begin{equation}
\label{eq:statement2}
|\partial_{\theta_i} G_n(\vec{\theta})|< C V^2L_i^{1+\frac{n}{2}}e^{-\frac{L_i}{4\xi'}}.
\end{equation}

\section{Excitation energy} 
\label{sec:gap}
So far we have only investigated the ground state properties. Here let us discuss what we can say about excitations.

\subsection{Energy expectation value of variational state}
Let us consider an operator $\hat{O}$ of the form $\hat{O}=\int d^d\vec{x}\,\hat{O}_{\vec{x}}$ with local operators $\hat{O}_{\vec{x}}$. 
We construct a variational state $|O\rangle=\delta\hat{O}|0\rangle$, which is orthogonal to the ground state by definition.  Its energy expectation value measured from the ground state energy is given by
\begin{eqnarray}
\Delta_O&\equiv&\frac{\langle O|\hat{H}|O\rangle}{\langle O|O\rangle}-E=\frac{\langle \delta\hat{O}^\dagger[\hat{H},\delta\hat{O}]\rangle}{\langle \delta\hat{O}^\dagger \delta\hat{O}\rangle}\notag\\
&=&\frac{\int d^d\vec{x}d^d\vec{y}\langle \delta\hat{O}_{\vec{x}}^\dagger [\hat{H},\delta\hat{O}_y]\rangle}{\int d^d\vec{x}d^d\vec{y}\langle \delta\hat{O}_{\vec{x}}^\dagger \delta\hat{O}_{\vec{y}}\rangle}.\label{energy}
\end{eqnarray}
The denominator is proportional to the system size $V=L_1L_2\ldots L_d$ because of the exponential decay of the correlation function. 
Similarly, the numerator is also proportional to $V$ since the commutator $[\hat{H},\delta\hat{O}_y]$ is still local owing to the locality of the Hamiltonian. Therefore, the energy expectation value $\Delta_O$ can be at most $O(V^0)$~\footnote{In order to achieve higher energy states whose excitation energy grows as $O(V^\epsilon)$ with $\epsilon>0$, one needs a non-local operation rather than simply superposing local perturbations.  In fact, when $\hat{O}=\hat{O}_1\hat{O}_2$ is a product of two well-separated local operators, we have $\Delta_O\simeq \Delta_{O_1}+\Delta_{O_2}$ and the correction decays exponentially with their distance. This implies that one can get a higher-energy state by creating many local excitations simultaneously.}.

We show that the excitation energy of locally excited states is almost independent of the flux $\theta$.  To this end, 
suppose that the Hamiltonian $\hat{H}[A]$ has a U(1) symmetry satisfying Eq.~\eqref{Htrans}.  
We assume the form $\hat{O}[A]=\int d^d\vec{x}\,\hat{O}_{\vec{x}}[A]$ with local operators $\hat{O}_{\vec{x}}[A]$ obeying in Eq.~\eqref{Htrans}.  We set $A(x)=\frac{\theta}{L}$ and construct the variational state $|O[\frac{\theta}{L}]\rangle=\delta\hat{O}[\frac{\theta}{L}]|\theta\rangle$.    Now, note that the last expression of Eq.~\eqref{energy} is written in terms of the expectation value of local operators. Thus we can apply the result in Sec.~\ref{sec:exp}. Therefore, the derivative $\partial_\theta\Delta_{O[\frac{\theta}{L}]}$ is bounded by $F_1$ in Eq.~\eqref{decay2} with $R=\frac{L}{4}$.  

\subsection{Insensitivity of excitation gap}
Now let us discuss the $\theta$-dependence of the true excitation gap. 
More precisely, here $\Delta_\theta$ denotes the gap to the first excited state $|1\rangle_\theta$ in the same sector of the conserved U(1) charge.  We assume that there exits a local operator $\hat{O}_0$ such that the state $\hat{O}_0|0\rangle$ has a nonzero overlap with $|1\rangle$, i.e., $|\langle1|\hat{O}_0|0\rangle|^2=w>0$. (The weight $w$ can be proportional to $L^{-\alpha}$ with $\alpha\geq0$. The expectation value of the excitation energy $\Delta_{O'}$ can be much larger than $\Delta$.)  Then, by applying the energy filter~\cite{Verstraete}, one can construct a low-energy local operator $\hat{O}$ from $\hat{O}_0$ such that, for any $\epsilon>0$, (i) the excitation energy $\Delta_O$ satisfies $\Delta\leq\Delta_O\leq\Delta(1+\epsilon)+\delta$, where $\delta=\frac{\tilde{C}}{w}(\tilde{R}/\xi_0)^\ell e^{-\epsilon\tilde{R}/\tilde{\xi}}$ is an exponentially small correction with some power $\ell$ and $\tilde{\xi}\equiv\frac{\sqrt{2}v}{\Delta}+\epsilon \xi_0$ and (ii) the support $\Omega$ of $\hat{O}$ is finite and includes the support of $\hat{O}_0$ inside. Here, $\tilde{R}=\text{dist}(\partial\Omega,\hat{O}_0)$ denotes the minimum distance between the boundary of $\Omega$ and the support of $\hat{O}_0$~\cite{Verstraete,Koma}. We reproduce the derivation in Appendix C.  

Using this operator $\hat{O}$, we prove that $\Delta_\theta$ does not depend much on $\theta$ for a large system size. Our argument is proof by contradiction.  
Suppose that the gap becomes smaller at $\theta=\theta_0$ ($0<\theta_0<2\pi$) than the value $\Delta_0$ at $\theta=0$. Namely, there exists $\xi$ ($0<\xi<1$) such that 
\begin{equation}
\Delta_{\theta_0}=\xi\Delta_{0}.\label{eq:assumption}
\end{equation}
By setting $\epsilon=\frac{1-\xi}{2\xi}$ and $\tilde{R}=\frac{L}{2}$, for example, we can construct a local operator $\hat{O}[\frac{\theta_0}{L}]$ such that 
\begin{equation}
\Delta_{\theta_0}\leq \Delta_{O[\frac{\theta}{L}]}\leq\Delta_{\theta_0}(1+\epsilon)+\delta=\tfrac{1+\xi}{2}\Delta_{0}+\delta.
\end{equation}
Since $\Delta_{O[\frac{\theta}{L}]}$ does not depend much on $\theta$ as proven above, it in turn implies that 
\begin{equation}
\Delta_{O[0]}<\tfrac{1+\xi}{2}\Delta_{0}+\delta+\delta'
\end{equation}
with another exponentially small correction $\delta'$. We can make $\delta+\delta'$ smaller than $\frac{1-\xi}{2}\Delta_{0}$ by choosing a sufficiently large $L$ so that 
\begin{equation}
\Delta_{O[0]}<\Delta_{0}
\end{equation}
This is a contradiction, since the energy expectation value of a variational state can never be smaller than the real excitation energy.
Therefore, the assumption in Eq.~\eqref{eq:assumption} must be wrong and $\Delta_{\theta_0}$ cannot be smaller than $\Delta_0$ by any finite amount. In fact, the excitation gap $\Delta$ can depend on $\theta$ at most by an exponentially small amount with respect to the system size.

This, in particular, indicates that that the excitation energy $\Delta_\theta$ never vanishes if $\Delta_0$ is finite in the limit of large system size.  This corollary completes, with one remaining assumption on the existence of the local operator $\hat{O}$, the proof of the higher-dimensional Lieb-Schultz-Mattis theorem by Oshikawa~\cite{Oshikawa2000}, without assuming the reality of the Hamiltonian.

\section{Concluding remarks}
\label{sec:conclude}
We demonstrated the $\theta$-independence of static responses among other things. In fact, one can replace $\hat{H}-E$ in Eqs.~\eqref{decay2}--\eqref{decay4} by $\hat{H}-E- \omega$ as long as $\omega<\Delta$, which simply gives the ``effective gap" $\Delta-\omega$.  Therefore, the dynamical susceptibility with a frequency lower than $\Delta$ can be covered by the method developed in this work.  

The $\theta$-dependence of the ground state energy is related to the transport properties: the first derivative $\partial_\theta E_\theta$  represents the persistent current and the second derivative gives the Drude weight via the Kohn formula $D=\frac{\pi L^2}{V}\partial_\theta^2 E_\theta$~\cite{Kohn,Scalapino,OshikawaDrude}. 
Our argument for expectation values and response properties proves that both of them are exponentially small with the system size in U(1) symmetric gapped phases.  

In the derivation we assumed the uniqueness of the ground state. However, similar statements should hold even when a finite (quasi-)degeneracy originates from spontaneous breaking of discrete symmetries or the presence of topological orders~\cite{Koma}. Let us denote by $\{|0_\alpha\rangle\}_{\alpha=1}^q$ the $q$-fold (quasi-)degenerate ground states.   In general, off-diagonal matrix elements $\langle0_\alpha|\hat{O}|0_\beta\rangle$ ($\alpha\neq\beta$) are expected to be exponentially small with the system size as long as the operator $\hat{O}=\sum_{\vec{x}}\hat{O}_{\vec{x}}$ is a sum (or integral) of local operators. They should be proportional to $e^{-\frac{V}{\xi^d}}$ in phases with discrete symmetry breaking and $e^{-\frac{L}{\xi}}$ for topologically ordered phases.  Assuming this scaling, the degenerate case does not seem fundamentally different, but we will leave the concrete analysis to future work.

\begin{acknowledgments}
H. W. thanks Tohru Koma for fruitful discussions and for explaining Ref.~\onlinecite{Koma} in detail.  This work is supported by JSPS KAKENHI Grant Numbers JP17K17678.
\end{acknowledgments}

\bibliography{references}

\clearpage
\appendix

\onecolumngrid
\section{Bound of two-point correlation functions}
In this appendix we prove the exponential decay of correlation functions in Eqs.~(1) and (2) of the main text.  The proof involves a few math formulas. For example, for $x\geq0$, we have
\begin{eqnarray}
&&0\leq\int_{0}^{x}\frac{dy}{2\pi}\frac{e^{y}-1}{y}<\frac{e^{x}}{x},\label{exp}\\
&&0<\text{erfc}(x)\equiv \frac{2}{\sqrt{\pi}}\int_x^\infty dy\,e^{-y^2}\leq e^{-x^2}.\label{erfc}
\end{eqnarray}

We will also use
\begin{eqnarray}
|\langle\hat{O}\,\hat{O}'\rangle|&\leq&\sqrt{\langle\hat{O}\hat{O}^\dagger\rangle\langle\hat{O}'^\dagger\hat{O}'\rangle}\leq\|\hat{O}\|\|\hat{O}'\|,\label{Schwartz}\\
\langle\hat{O}^\dagger f(\hat{H})\hat{O}\rangle&\leq& \langle\hat{O}^\dagger \hat{O}\rangle f(\Delta)\quad\text{for a positive and monotonically decreasing function $f(\mathcal{E})$}\label{Complete},\\
\langle\hat{O}\hat{O}'(\tau)\rangle&=&e^{-\alpha t^2}\left[\langle\hat{O}\hat{O}'(\tau)\rangle (e^{\alpha t^2}-e^{-\alpha \tau^2})+\langle\hat{O}\hat{O}'(\tau)\rangle e^{-\alpha \tau^2}\right]\notag\\
&=&e^{-\alpha t^2}\left[\langle\hat{O}\hat{O}'(\tau)\rangle (e^{\alpha t^2}-e^{-\alpha \tau^2})+\langle\hat{O}'(\tau)\hat{O}\rangle e^{-\alpha \tau^2}+\langle[\hat{O},\hat{O}'(\tau)]\rangle e^{-\alpha \tau^2}\right].\label{split3}
\end{eqnarray}
Here, Eq.~\eqref{Schwartz} is the Schwartz inequality and Eq.~\eqref{split3} follows just by the definition of the commutation relation.

The following mathematical identities are valid for arbitrary $\mathcal{E}, \alpha, t>0$:
\begin{eqnarray}
F_+(\mathcal{E})&\equiv&\int_{-\infty}^\infty\frac{d\tau}{2\pi i}\frac{e^{+i\mathcal{E}\tau}}{\tau-it}(e^{\alpha t^2}-e^{-\alpha \tau^2})=\frac{1}{2\sqrt{\pi \alpha}}\int_{0}^\infty d\omega\,e^{+\omega t-\frac{(\mathcal{E}+\omega)^2}{4\alpha}}=\frac{1}{2}e^{\alpha t^2-t\mathcal{E}}\text{erfc}\left(\frac{\mathcal{E}-2\alpha t}{2\sqrt{\alpha}}\right)>0,\label{Identity1}\\
F_-(\mathcal{E})&\equiv&\int_{-\infty}^\infty\frac{d\tau}{2\pi i}\frac{e^{-i\mathcal{E}\tau}}{\tau-it}e^{-\alpha \tau^2}=\frac{1}{2\sqrt{\pi \alpha}}\int_{0}^\infty d\omega\,e^{-\omega t-\frac{(\mathcal{E}+\omega)^2}{4\alpha}}=\frac{1}{2}e^{\alpha t^2+t\mathcal{E}}\text{erfc}\left(\frac{\mathcal{E}+2\alpha t}{2\sqrt{\alpha}}\right)>0.\label{Identity2}
\end{eqnarray}
Using the property Eq.~\eqref{erfc}, we have
\begin{eqnarray}
0<F_{\pm}(\mathcal{E})\leq\frac{1}{2}e^{-\frac{E^2}{4\alpha}}\quad \text{when}\quad 0<t\leq\frac{\mathcal{E}}{2\alpha}.
\end{eqnarray}

Finally, the Lieb-Robinson bound will be used to derive an upper bound of commutation relations:
\begin{eqnarray}
\|[\hat{O},\hat{V}(t)]\|\leq C_{OV}e^{-\frac{R}{\xi_0}}(e^{\frac{v|t|}{\xi_0}}-1).\label{LRbound}
\end{eqnarray}
Here, $\xi_0$ and $v$ are constants, independent of the choice of $\hat{O}$ and $\hat{V}$. 

\subsection{Correlation function $F_{0}$}
Let us start with $F_0\equiv\langle\delta\hat{O}\,\delta\hat{V}\rangle$.   Instead of directly dealing with $F_{0}$, here we evaluate
\begin{equation}
F_0(t)\equiv\langle\delta\hat{O}\,\delta\hat{V}(it)\rangle=\langle\delta\hat{O}\,e^{-t(\hat{H}-E)}\delta\hat{V}\rangle.\label{app:f0}
\end{equation}
Using the complex analysis, we can express $\langle\delta\hat{O}\,\delta\hat{V}(it)\rangle$ in the form of the integral
\begin{eqnarray}
F_0(t)=\oint\frac{dz}{2\pi i}\frac{\langle\delta\hat{O}\,\delta\hat{V}(z)\rangle}{z-it}=\lim_{S\rightarrow\infty}\int_{-S}^S\frac{d\tau}{2\pi i}\frac{\langle\delta\hat{O}\,\delta\hat{V}(\tau)\rangle}{\tau-it}+\lim_{S\rightarrow\infty}\int_{0}^{\pi}\frac{d\phi}{2\pi}Se^{i\phi}\frac{\langle\delta\hat{O}\,\delta\hat{V}(Se^{i\phi})\rangle}{Se^{i\phi}-it}.\label{split11}
\end{eqnarray}
The second integral in the right-hand side of Eq.~\eqref{split11} vanishes in the limit of $S\rightarrow \infty$:
\begin{eqnarray}
\int_{0}^{\pi}\frac{d\phi}{2\pi}\left|\frac{\langle\delta\hat{O}\,\delta\hat{V}(Se^{i\phi})\rangle}{1-ie^{-i\phi} t/S}\right|\leq \|\delta\hat{O}\|\|\delta\hat{V}\| \int_{0}^{\pi}\frac{d\phi}{2\pi}\frac{e^{-S\Delta\sin\phi}}{\sqrt{(1-\frac{t}{S}\sin\phi)^2+(\frac{t}{S}\cos\phi)^2}}\leq \|\delta\hat{O}\|\|\delta\hat{V}\|\frac{1-e^{-S\Delta}}{2(S-t)\Delta}\rightarrow0.
\end{eqnarray}
We used Eqs.~\eqref{Schwartz} and \eqref{Complete} in the first step.

The remaining integral in Eq.~\eqref{split11} can be split into four using Eq.~\eqref{split3}:
\begin{equation}
F_0(t)=\int_{-\infty}^\infty\frac{d\tau}{2\pi i}\frac{\langle\delta\hat{O}\,\delta\hat{V}(\tau)\rangle}{\tau-it}=e^{-\alpha t^2}(I_1+I_2+I_3+I_4),\label{I123}
\end{equation}
where
\begin{eqnarray}
I_1&\equiv&\int_{-\infty}^\infty\frac{d\tau}{2\pi i}\frac{\langle\delta\hat{O}\,\delta\hat{V}(\tau)\rangle}{\tau-it}(e^{\alpha t^2}-e^{-\alpha \tau^2}),\\
I_2&\equiv&\int_{-\infty}^\infty\frac{d\tau}{2\pi i}\frac{\langle\delta\hat{V}(\tau)\,\delta\hat{O}\rangle}{\tau-it}e^{-\alpha \tau^2},\\
I_3&\equiv&\int_{|\tau|>T}\frac{d\tau}{2\pi i}\frac{\langle[\hat{O},\hat{V}(\tau)]\rangle}{\tau-it}e^{-\alpha \tau^2},\\
I_4&\equiv&\int_{-T}^T\frac{d\tau}{2\pi i}\frac{\langle[\hat{O},\hat{V}(\tau)]\rangle}{\tau-it}e^{-\alpha \tau^2}.
\end{eqnarray}
The parameters $T$ and $\alpha$ are chosen as
\begin{equation}
T\equiv\frac{2R}{\xi\Delta}, \quad\alpha\equiv\frac{\Delta^2\xi}{4R} ,\quad R\equiv\text{dist}(\hat{O},\hat{V}),
\end{equation}
so that $\frac{\Delta^2}{4\alpha}=\alpha T^2=\frac{R}{\xi}\gg1$.

Integrals $I_1$ and $I_2$ can be performed with the help of the identities in Eq.~\eqref{Identity1} and~\eqref{Identity2}: 
\begin{eqnarray}
I_1&=&\langle\delta\hat{O}\,F_+(\hat{H}-E)\,\delta\hat{V}\rangle,\\
I_2&=&\langle\delta\hat{V}\,F_-(\hat{H}-E)\,\delta\hat{O}\rangle.
\end{eqnarray}
Then, assuming $0<t\leq\frac{\Delta}{2\alpha}$ and using Eqs.~\eqref{Schwartz} and \eqref{Complete}, we get
\begin{eqnarray}
|I_1|&\leq&\sqrt{\langle\delta\hat{O}\delta\hat{O}^\dagger\rangle\langle\delta\hat{V}^\dagger F_+(\hat{H}-E)^2\delta\hat{V}\rangle}\leq\frac{1}{2}\sqrt{\langle\delta\hat{O}\delta\hat{O}^\dagger\rangle\langle\delta\hat{V}^\dagger e^{-\frac{(\hat{H}-E)^2}{2\alpha}}\delta\hat{V}\rangle}\leq\frac{\|\delta\hat{O}\|\|\delta\hat{V}\|}{2}e^{-\frac{\Delta^2}{4\alpha}},\label{estimateI1}\\
|I_2|&\leq&\sqrt{\langle\delta\hat{V}\delta\hat{V}^\dagger\rangle\langle\delta\hat{O}^\dagger F_-(\hat{H}-E)^2\delta\hat{O}\rangle}\leq\frac{1}{2}\sqrt{\langle\delta\hat{V}\delta\hat{V}^\dagger\rangle\langle\delta\hat{O}^\dagger e^{-\frac{(\hat{H}-E)^2}{2\alpha}}\delta\hat{O}\rangle}\leq\frac{\|\delta\hat{O}\|\|\delta\hat{V}\|}{2}e^{-\frac{\Delta^2}{4\alpha}}.\label{estimateI2}
\end{eqnarray}
The integral $I_3$ can be bounded by Eqs.~\eqref{Schwartz} and \eqref{erfc}:
\begin{eqnarray}
|I_3|&\leq&\int_{|\tau|>T}\frac{d\tau}{2\pi}\frac{|\langle[\delta\hat{O},\delta\hat{V}(\tau)]\rangle|}{\sqrt{\tau^2+t^2}}e^{-\alpha \tau^2}< 4\|\delta\hat{O}\|\|\delta\hat{V}\|\int_{T}^{\infty}\frac{d\tau}{2\pi}\frac{e^{-\alpha \tau^2}}{T}\leq\frac{\|\delta\hat{O}\|\|\delta\hat{V}\|}{\sqrt{\pi\alpha T^2}}e^{-\alpha T^2}.\label{estimateI3}
\end{eqnarray}
Finally, the integral $I_4$ can be bounded by Lieb-Robinson bound Eq.~\eqref{LRbound} and the inequality in Eq.~\eqref{exp}:
\begin{eqnarray}
|I_4|&\leq&\int_{|\tau|<T}\frac{d\tau}{2\pi}\frac{|\langle[\hat{O},\hat{V}(\tau)]\rangle|}{\sqrt{\tau^2+t^2}}e^{-\alpha \tau^2}\leq2C_{OV}e^{-\frac{R}{\xi_0}}\int_{0}^T\frac{d\tau}{2\pi}\frac{e^{\frac{v\tau}{\xi_0}}-1}{\tau}\leq2C_{OV}\frac{e^{\frac{vT-R}{\xi_0}}}{\frac{vT}{\xi_0}}=\frac{\xi_0\Delta}{v}\frac{C_{OV}}{R/\xi}e^{-\frac{\xi-\frac{2v}{\Delta}}{\xi_0}\frac{R}{\xi}},
\end{eqnarray}

All in all, when $\frac{2R}{\xi}\geq  t\Delta>0$, we have
\begin{eqnarray}
|F_0(t)|&\leq& e^{-\alpha t^2}(|I_1|+|I_2|+|I_3|+|I_4|)\notag\\
&\leq&e^{-\frac{4\xi}{R}(t\Delta)^2}\|\delta\hat{O}\|\|\delta\hat{V}\|\left(e^{-\frac{R}{\xi}}+\tfrac{1}{\sqrt{\pi R/\xi}}e^{-\frac{R}{\xi}}+\frac{\xi_0\Delta}{v}\frac{C_{OV}}{\|\delta\hat{O}\|\|\delta\hat{V}\|\, R/\xi}e^{-\frac{\xi-\frac{2v}{\Delta}}{\xi_0}\frac{R}{\xi}}\right).
\end{eqnarray}
If we set $\xi=\xi_0+\frac{2v}{\Delta}$, all terms have a factor $e^{-\frac{R}{\xi}}$. When $t\Delta\geq\frac{2R}{\xi}$, we can directly evaluate $F_0(t)$ in Eq.~\eqref{app:f0} using Eqs.~\eqref{Schwartz} and \eqref{Complete}.  At the end, we have
\begin{eqnarray}
|F_0(t)|&\leq&
\begin{cases}
\|\delta\hat{O}\|\|\delta\hat{V}\|\left(
1+\frac{1}{\sqrt{\pi R/\xi}}+\frac{2\xi_0}{\xi-\xi_0}\frac{C_{OV}}{\|\delta\hat{O}\|\|\delta\hat{V}\|\,R/\xi}
\right)e^{-\frac{4\xi}{R}(t\Delta)^2}e^{-\frac{R}{\xi}}&\left(\frac{2R}{\xi}\geq t\Delta>0\right)\\
\|\delta\hat{O}\|\|\delta\hat{V}\|e^{-\frac{2R}{\xi}}&\left(t\Delta\geq\frac{2R}{\xi}\right).
\end{cases}\label{app:result1}
\end{eqnarray}
The equal-time correlation $F_{0}=\langle\delta\hat{O}\,\delta\hat{V}\rangle$ is the limit of $t\rightarrow+0$.

\subsection{Correlation function $F_{n}$}
\label{appFn}
Next let us discuss the correlation function with $(\hat{H}-E)^{-n}$:
\begin{eqnarray}
F_{n}&\equiv&\langle\delta\hat{O}\frac{1}{(\hat{H}-E)^n}\delta\hat{V}\rangle=\int_{0}^\infty dt\,\frac{t^{n-1}}{(n-1)!}\langle\delta\hat{O}\,e^{-t(\hat{H}-E)}\delta\hat{V}\rangle=\int_{0}^\infty dt\,\frac{t^{n-1}}{(n-1)!}F_0(t)=I_4+I_5,\\
I_4&\equiv&\int_{T}^\infty dt\,\frac{t^{n-1}}{(n-1)!}\langle\delta\hat{O}\,e^{-t(\hat{H}-E)}\delta\hat{V}\rangle,\\
I_5&\equiv&\int_{0}^T dt\,\frac{t^{n-1}}{(n-1)!}f_0(t).
\end{eqnarray}
The integral $I_4$ can be estimated by Eqs.~\eqref{Schwartz} and \eqref{Complete}:
\begin{eqnarray}
|I_4|&\leq& \|\delta\hat{O}\|\|\delta\hat{V}\| \int_{T}^\infty dt\,\,\frac{t^{n-1}}{(n-1)!}e^{-t\Delta}=\frac{\|\delta\hat{O}\|\|\delta\hat{V}\|}{\Delta^{n}}\sum_{m=0}^{n-1}\frac{(T\Delta)^m}{m!}e^{-T\Delta}=\frac{\|\delta\hat{O}\|\|\delta\hat{V}\|}{\Delta^{n}}\sum_{m=0}^{n-1}\frac{1}{m!}\left(\frac{2R}{\xi}\right)^me^{-\frac{2R}{\xi}}.
\end{eqnarray}
For the integral $I_5$, we can use the first line of Eq.~\eqref{app:result1}. Writing $c_n\equiv\int_0^{\infty}dx\,\frac{x^{n-1}}{(n-1)!}e^{-x^2}$, we have $\int_0^Tdt\,\frac{t^{n-1}}{(n-1)!}e^{-\frac{4\xi}{R}(t\Delta)^2}\leq \frac{c_n}{\Delta^n}\left(\frac{4R}{\xi}\right)^{n/2}$ and
\begin{eqnarray}
|I_5|\leq\frac{c_n}{\Delta^n}\left(\frac{4R}{\xi}\right)^{n/2}\|\delta\hat{O}\|\|\delta\hat{V}\|\left(1+\frac{1}{\sqrt{\pi R/\xi}}+\frac{2\xi_0}{\xi-\xi_0}\frac{C_{OV}}{\|\delta\hat{O}\|\|\delta\hat{V}\|\,R/\xi}\right)e^{-\frac{R}{\xi}}.
\end{eqnarray}
Therefore, $|F_n|\leq|I_4|+|I_5|$ is exponentially suppressed.  For a sufficiently large $R/\xi\gg1$, the dominant contribution to $F_{n}$ comes from the first term in $|I_5|$.

\section{Three point correlation functions}
Here we derive the bound in Eqs. (5) and (6) in the main text. To this end, we evaluate the correlation function of the following form:
\begin{eqnarray}
G(s,t)&\equiv&\langle\delta\hat{a}\,e^{-s(\hat{H}-E)}\delta\hat{b}\,e^{-t(\hat{H}-E)}\delta\hat{c}\rangle=\langle\delta\hat{a}\,e^{-s(\hat{H}-E)}\delta\hat{b}\,\delta\hat{c}(it)\rangle.
\end{eqnarray}
for $s,t\in(0,T)$ with $T=\frac{2R}{\xi\Delta}$.  Later we will set ``$\hat{a}=\hat{V}$, $\hat{b}=\hat{O}$, and $\hat{c}=\hat{O}'$" or ``$\hat{a}=\hat{O}$, $\hat{b}=\hat{V}$, and $\hat{c}=\hat{O}'$" with $R\equiv\text{min}(\text{dist}(\hat{O},\hat{V}),\text{dist}(\hat{O'},\hat{V}))\gg\xi$.  Once $|G(s,t)|$ is bounded, then the correlation functions in Eqs. (3) and (4) in the main text can be evaluated by performing the integral $\int ds s^{m-1}\int dt t^{n-1}G(s,t)$ as we did in Sec.~\ref{appFn}.

As before, we split the integral into those pieces which we know how to estimate:
\begin{eqnarray}
G(s,t)&=&\int_{-\infty}^\infty\frac{d\tau}{2\pi i}\frac{\langle\delta\hat{a}\,e^{-s(\hat{H}-E)}\delta\hat{b}\,\delta\hat{c}(\tau)\rangle}{\tau-it}=e^{-\alpha t^2}(I_1+I_2+I_3),
\end{eqnarray}
where $\alpha=\frac{\Delta^2\xi}{4R}$ and 
\begin{eqnarray}
I_1&\equiv&\int_{-\infty}^\infty\frac{d\tau}{2\pi i}\frac{\langle\delta\hat{a}\,e^{-s(\hat{H}-E)}\delta\hat{b}\,\delta\hat{c}(\tau)\rangle}{\tau-it}(e^{\alpha t^2}-e^{-\alpha \tau^2}),\\
I_2&\equiv&\int_{|\tau|>T}\frac{d\tau}{2\pi i}\frac{\langle\delta\hat{a}\,e^{-s(\hat{H}-E)}\delta\hat{b}\,\delta\hat{c}(\tau)\rangle}{\tau-it}e^{-\alpha \tau^2},\\
I_3&\equiv&\int_{-T}^T\frac{d\tau}{2\pi i}\frac{\langle\delta\hat{a}\,e^{-s(\hat{H}-E)}\delta\hat{b}\,\delta\hat{c}(\tau)\rangle}{\tau-it}e^{-\alpha \tau^2}=\int_{-T}^T\frac{d\tau}{2\pi i}\frac{e^{-\alpha \tau^2}}{\tau-it}\langle\delta\hat{a}(-is)\delta\hat{b}\,\delta\hat{c}(\tau)\rangle\notag\\
&=&\int_{-T}^T\frac{d\tau}{2\pi i}\frac{e^{-\alpha \tau^2}}{\tau-it}\int_{-\infty}^\infty\frac{d\sigma}{2\pi i}\frac{\langle\delta\hat{a}(\sigma)\delta\hat{b}\,\delta\hat{c}(\tau)\rangle}{\sigma+is}\notag\\
&=&\int_{-T}^T\frac{d\tau}{2\pi i}\frac{e^{-\alpha \tau^2}}{\tau-it}e^{-\alpha s^2}(I_{31}+I_{32}+I_{33}+I_{34}),
\end{eqnarray}
and
\begin{eqnarray}
I_{31}&\equiv&\int_{-\infty}^\infty\frac{d\sigma}{2\pi i}\frac{\langle\delta\hat{a}(\sigma)\delta\hat{b}\,\delta\hat{c}(\tau)\rangle}{\sigma+is}(e^{\alpha s^2}-e^{-\alpha \sigma^2}),\\
I_{32}&\equiv&\int_{-\infty}^\infty\frac{d\sigma}{2\pi i}\frac{\langle\delta\hat{b}\,\delta\hat{c}(\tau)\delta\hat{a}(\sigma)\rangle}{\sigma+is}e^{-\alpha \sigma^2},\\
I_{33}&\equiv&\int_{|\tau|>T}\frac{d\sigma}{2\pi i}\frac{\langle[\delta\hat{a}(\sigma),\delta\hat{b}\,\delta\hat{c}(\tau)]\rangle}{\sigma+is}e^{-\alpha \sigma^2},\\
I_{34}&\equiv&\int_{-T}^T\frac{d\sigma}{2\pi i}\frac{\langle[\delta\hat{a}(\sigma),\delta\hat{b}\,\delta\hat{c}(\tau)]\rangle}{\sigma+is}e^{-\alpha \sigma^2}.
\end{eqnarray}
In the same way as Eqs.~\eqref{estimateI1} and~\eqref{estimateI2}, we have
\begin{eqnarray}
|I_1|,|I_{31}|,|I_{32}|\leq\frac{\|\delta\hat{a}\|\|\delta\hat{b}\|\|\delta\hat{c}\|}{2}e^{-\frac{R}{\xi}}.
\end{eqnarray}
Following Eq.~\eqref{estimateI3}, we get
\begin{eqnarray}
|I_2|, |I_{33}|\leq\frac{\|\delta\hat{a}\|\|\delta\hat{b}\|\|\delta\hat{c}\|}{\sqrt{\pi R/\xi}}e^{-\frac{R}{\xi}}.
\end{eqnarray}

Therefore it remains to estimate $I_{34}$:
\begin{eqnarray}
I_{34}&=&\int_{-T}^T\frac{d\sigma}{2\pi i}\frac{\langle\delta\hat{b}\,[\hat{a}(\sigma),\hat{c}(\tau)]\rangle+\langle[\hat{a}(\sigma),\hat{b}]\,\delta\hat{c}(\tau)\rangle}{\sigma+is}e^{-\alpha \sigma^2}.
\end{eqnarray}

\subsection{When $\hat{a}=\hat{V}$, $\hat{b}=\hat{O}$, and $\hat{c}=\hat{O}'$}
In this case we can simply apply the Lieb-Robinson bound Eq.~\eqref{LRbound}:
\begin{eqnarray}
|I_{34}|&\leq&\int_{-T}^T\frac{d\sigma}{2\pi}\frac{e^{-\alpha \sigma^2}}{\sqrt{\sigma^2+s^2}}(|\langle\delta\hat{O}\,[\hat{V}(\sigma),\hat{O}'(\tau)]\rangle|+|\langle[\hat{V}(\sigma),\hat{O}]\,\delta\hat{O}'(\tau)\rangle|)\notag\\
&\leq&\int_{-T}^T\frac{d\sigma}{2\pi}\frac{e^{-\alpha \sigma^2}}{\sqrt{\sigma^2+s^2}}(\|\delta\hat{O}\|\,C_{VO'}e^{\frac{v(|\sigma|+|\tau|)-|\vec{x}_{c}-\vec{x}_{a}|}{\xi_0}}+\|\delta\hat{O}'\|\,C_{VO}e^{\frac{v|\tau|-|\vec{x}_{b}-\vec{x}_{a}|}{\xi_0}})\notag\\
&\leq&\int_{-T}^T\frac{d\sigma}{2\pi}\frac{e^{-\alpha \sigma^2}}{\sqrt{\sigma^2+s^2}}(\|\delta\hat{O}\|\,C_{VO'}+\|\delta\hat{O}'\|\,C_{VO})e^{\frac{2vT-R}{\xi_0}}\notag\\
&\leq& F(s)(\|\delta\hat{O}\|\,C_{VO'}+\|\delta\hat{O}'\|\,C_{VO})e^{\frac{2vT-R}{\xi_0}},
\end{eqnarray}
where
\begin{equation}
F(x)\equiv\int_{-\infty}^\infty\frac{dy}{2\pi}\frac{e^{-\alpha y^2}}{\sqrt{x^2+y^2}}.
\end{equation}
Collecting all terms and setting $\xi'=\xi_0+\frac{4v}{\Delta}$, we get
\begin{eqnarray}
\frac{|G(s,t)|}{\|\delta\hat{V}\|\|\delta\hat{O}\|\|\delta\hat{O}'\|}\leq\Big(\tfrac{1}{2}+\tfrac{1}{\sqrt{\pi R/\xi}}\Big)e^{-\alpha t^2-\frac{R}{\xi}}+F(t)\left(1+\tfrac{1}{\sqrt{\pi R/\xi}}+F(s)\tfrac{\|\delta\hat{O}\|\,C_{VO'}+\|\delta\hat{O}'\|\,C_{VO}}{\|\delta\hat{V}\|\|\delta\hat{O}\|\|\delta\hat{O}'\|}\right)e^{-\alpha (s^2+t^2)-\frac{R}{\xi}}.
\end{eqnarray}
The function $F(x)$ itself may diverge at $x=0$, but it only appears in the following integral at the end:
\begin{equation}
\int_{0}^\infty dx\,\frac{x^{n-1}}{(n-1)!}e^{-\alpha x^2}F(x)=\int_{0}^\infty dx\int_{-\infty}^\infty\frac{dy}{2\pi}\frac{x^{n-1}}{(n-1)!}\frac{e^{-\alpha (x^2+y^2)}}{\sqrt{x^2+y^2}}\leq\frac{1}{2}\int_{0}^\infty dr \,\frac{r^{n-1}}{(n-1)!}e^{-\alpha r^2}=\frac{c_n}{2\alpha^{n/2}}.
\end{equation}

\subsection{When $\hat{a}=\hat{O}$, $\hat{b}=\hat{V}$, and $\hat{c}=\hat{O}'$}
This case requires a new relation:
\begin{eqnarray}
|\langle\delta\hat{O}(\sigma_1)\,\delta\hat{O}'(\sigma_2)\,\delta\hat{V}\rangle|&\leq& \|\delta\hat{V}\|\|\delta\hat{O}\|\|\delta\hat{O}'\|\Big(1+\tfrac{1}{\sqrt{\pi R/\xi}}+\tfrac{\|\delta\hat{O}\|C_{O',[H,V]}+\|\delta\hat{O}'\|C_{O,[H,V]}}{ \|\delta\hat{V}\|\|\delta\hat{O}\|\|\delta\hat{O}'\|\Delta}\sqrt{\tfrac{R}{\pi\xi}}\Big)e^{-\frac{R}{\xi}}\notag\\
&\equiv& B(R/\xi)e^{-\frac{R}{\xi}}\label{appnew}
\end{eqnarray}
for $|\sigma_1|,|\sigma_2|\leq T$. Given this, we can get
\begin{eqnarray}
|I_{34}|&\leq&\int_{-T}^T\frac{d\sigma}{2\pi}\frac{e^{-\alpha \sigma^2}}{\sqrt{\sigma^2+s^2}}(|\langle\delta\hat{V}\,[\hat{O}(\sigma),\hat{O}'(\tau)]\rangle|+|\langle[\hat{O}(\sigma),\hat{V}]\,\delta\hat{O}'(\tau)\rangle|)\notag\\
&\leq&\int_{-T}^T\frac{d\sigma}{2\pi}\frac{e^{-\alpha \sigma^2}}{\sqrt{\sigma^2+s^2}}\Big[2B(R/\xi)e^{-\frac{R}{\xi}}+\|\delta\hat{O}'\|\,C_{OV}e^{\frac{vT-R}{\xi_0}}\Big]\notag\\
&\leq& F(s)\Big[2B(R/\xi)+\|\delta\hat{O}'\|\,C_{OV}\Big]e^{-\frac{R}{\xi}}.
\end{eqnarray}

The bound Eq.~\eqref{appnew} can be verified in the following way. Again using Eq.~\eqref{split3}, we have
\begin{eqnarray}
\langle\delta\hat{O}(\sigma_1)\,\delta\hat{O}'(\sigma_2)\,\delta\hat{V}\rangle=\lim_{t\rightarrow+0}\int_{-\infty}^{\infty}\frac{d\tau}{2\pi i}\frac{\langle\delta\hat{O}(\sigma_1)\,\delta\hat{O}'(\sigma_2)\,\delta\hat{V}(\tau)\rangle}{\tau-it}=I_1'+I_2'+I_3'+I_4',
\end{eqnarray}
where
\begin{eqnarray}
&&I_1'=\lim_{t\rightarrow+0}\int_{-\infty}^{\infty}\frac{d\tau}{2\pi i}\frac{\langle\delta\hat{O}(\sigma_1)\,\delta\hat{O}'(\sigma_2)\,\delta\hat{V}(\tau)\rangle}{\tau-it}(e^{\alpha t^2}-e^{-\alpha \tau^2}),\\
&&I_2'=\lim_{t\rightarrow+0}\int_{-\infty}^{\infty}\frac{d\tau}{2\pi i}\frac{\langle\delta\hat{V}(\tau)\,\delta\hat{O}(\sigma_1)\,\delta\hat{O}'(\sigma_2)\rangle}{\tau-it}e^{-\alpha \tau^2},\\
&&I_3'=\lim_{t\rightarrow+0}\int_{|\tau|>T}\frac{d\tau}{2\pi i}\frac{\langle[\delta\hat{O}(\sigma_1)\,\delta\hat{O}'(\sigma_2),\delta\hat{V}(\tau)]\rangle}{\tau-it}e^{-\alpha \tau^2},\\
&&I_4'=\lim_{t\rightarrow+0}\int_{-T}^{T}\frac{d\tau}{2\pi i}\frac{\langle[\delta\hat{O}(\sigma_1)\,\delta\hat{O}'(\sigma_2),\delta\hat{V}(\tau)]\rangle}{\tau-it}e^{-\alpha \tau^2}=\int_0^{T}\frac{d\tau}{2\pi i}\frac{e^{-\alpha \tau^2}}{\tau}\int_{-\tau} ^\tau du\langle[\delta\hat{O}(\sigma_1)\,\delta\hat{O}'(\sigma_2),\partial_u\hat{V}(u)]\rangle.
\end{eqnarray}
$I_1'$, $I_2'$,  $I_3'$ can be bounded in the same way as in Eqs.~\eqref{estimateI1}, \eqref{estimateI2} and \eqref{estimateI3}:
\begin{eqnarray}
|I_1|,|I_2|\leq\frac{\|\delta\hat{V}\|\|\delta\hat{O}\|\|\delta\hat{O}'\|}{2}e^{-\frac{R}{\xi}},\quad  |I_3|\leq\frac{\|\delta\hat{V}\|\|\delta\hat{O}\|\|\delta\hat{O}'\|}{\sqrt{\pi R/\xi}}e^{-\frac{R}{\xi}}.
\end{eqnarray}
For $I_4'$, we have
\begin{eqnarray}
|I_4'|&\leq&\int_0^{T}\frac{d\tau}{2\pi}\frac{e^{-\alpha \tau^2}}{\tau}\int_{-\tau} ^\tau du\big(|\langle\delta\hat{O}(\sigma_1)[\hat{O}'(\sigma_2),[\hat{H},\hat{V}(u)]]\rangle|+|\langle[\hat{O}(\sigma_1),[\hat{H},\hat{V}(u)]]\delta\hat{O}'(\sigma_2)\rangle|\big)\notag\\
&\leq&\int_0^{T}\frac{d\tau}{2\pi}\frac{e^{-\alpha \tau^2}}{\tau}\int_{-\tau} ^\tau du\big(\|\delta\hat{O}\|C_{O',[H,V]}+\|\delta\hat{O}'\|C_{O,[H,V]}\big)e^{\frac{2vT-R}{\xi_0}}\notag\\
&=&\frac{\|\delta\hat{O}\|C_{O',[H,O]}+\|\delta\hat{O}'\|C_{O,[H,V]}}{\Delta}\sqrt{\frac{R}{\pi\xi}}e^{\frac{2vT-R}{\xi_0}}.
\end{eqnarray}
In the derivation, we used the Lieb-Robinson bound Eq.~\eqref{LRbound} and $|u-\sigma_i|\leq2T$.

\section{Construction of the local operator approximately creating $|1\rangle$}
Here we discuss the construction of $\hat{O}$ starting from $\hat{O}_0$ defined in the main text.  Let $|1\rangle$ be the first excited state with the energy $\Delta=E_1-E_0$.  Suppose that the state $\hat{O}_0|0\rangle$ has a nonzero overlap with $|1\rangle$, i.e., $|\langle1|\hat{O}_0|0\rangle|^2=w>0$. In order to extract only the $|1\rangle$-component, let us apply the energy filter
\begin{eqnarray}
\hat{O}=\sqrt{\frac{\beta}{\pi}}\int_{-\infty}^{+\infty} dt\,\hat{O}_0(t)e^{-it\Delta-\beta t^2}=\sqrt{\frac{\beta}{\pi}}\int_{-\infty}^{+\infty} dt\,e^{it\hat{H}}\hat{O}_0e^{-it(\hat{H}+\Delta)-\beta t^2}.
\end{eqnarray}
with $\beta=\epsilon^2\frac{\Delta^2\xi_\epsilon}{2R}$.  Let us define two projection operators $\hat{Q}_{\text{low}}$ and $\hat{Q}_{\text{high}}$ onto energy windows $E_{\text{low}}\in [E_1,E_1+\epsilon\Delta]$ and $E_{\text{high}}\in (E_1+\epsilon\Delta,+\infty)$, respectively. We have
\begin{eqnarray}
\langle\hat{O}^\dagger\hat{Q}_{\text{high}}\hat{O}\rangle&=&\langle\hat{O}_0^\dagger e^{-\frac{1}{4\beta}(\hat{H}-E_1)^2}\hat{Q}_{\text{high}}e^{-\frac{1}{4\beta}(\hat{H}-E_1)^2}\hat{O}_0\rangle\leq \|\hat{O}_0\|^2 e^{-\epsilon^2\frac{\Delta^2}{2\beta}}=\|\hat{O}_0\|^2 e^{-\frac{R}{\xi_\epsilon}},\\
\langle\hat{O}^\dagger\hat{Q}_{\text{low}}\hat{O}\rangle&\geq& \langle\hat{O}^\dagger|1\rangle\langle1|\hat{O}\rangle=\langle\hat{O}_0^\dagger |1\rangle\langle1|\hat{O}_0\rangle= w.
\end{eqnarray}

Next, we want to approximate $\hat{O}$ by a local operator
\begin{eqnarray}
\hat{O}&=&\sqrt{\frac{\beta}{\pi}}\int_{-S}^{+S} dt\,e^{it\hat{H}_\Omega}\hat{O}_0e^{-it(\hat{H}_\Omega+\Delta)-\beta t^2}.
\end{eqnarray}
Here, $\Omega$ is a region including the support of $\hat{O}_0$, and $\hat{H}_\Omega$ denotes the Hamiltonian restricted onto the region.  Let us denote by $R$ the distance between $\partial\Omega$ and the support of $\hat{O}_0$.  We have
\begin{eqnarray}
&&\hat{O}-\hat{O}\notag=\sqrt{\frac{\beta}{\pi}}\int_{|t|>S}dte^{it\hat{H}}\hat{O}_0e^{-it(\hat{H}+\Delta)-\beta t^2}+\sqrt{\frac{\beta}{\pi}}\int_{-S}^{S} dt\int_0^tds\frac{d}{ds}(e^{is\hat{H}+i(t-s)\hat{H}_\Omega}\hat{O}_0e^{-i(t-s)\hat{H}_\Omega-is\hat{H}-it\Delta-\beta t^2})\notag\\
&=&\sqrt{\frac{\beta}{\pi}}\int_{|t|>S}dte^{it\hat{H}}\hat{O}_0e^{-it(\hat{H}+\Delta)-\beta t^2}+\sqrt{\frac{\beta}{\pi}}\int_{-S}^{S} dt\int_0^tds\,e^{is\hat{H}}[\hat{H}-\hat{H}_\Omega,e^{i(t-s)\hat{H}_\Omega}\hat{O}_0e^{-i(t-s)\hat{H}_\Omega}]e^{-is\hat{H}-it\Delta-\beta t^2}.
\end{eqnarray}
Using the Lieb-Robinson bound and setting $S=\frac{\sqrt{2}R}{\epsilon\xi_\epsilon\Delta}$, $\beta=\epsilon^2\frac{\Delta^2\xi_\epsilon}{2R}$, $\xi_\epsilon=\xi_0+\frac{\sqrt{2}v}{\epsilon\Delta}$, we get
\begin{eqnarray}
\|\hat{O}-\hat{O}\|
&\leq&\|\hat{O}_0\|\sqrt{\frac{\beta}{\pi}}\int_{|t|>S}dt\,e^{-\beta t^2}+C_{H_{\partial\Omega} O_0}Se^{\frac{vS-R}{\xi_0}}\sqrt{\frac{\beta}{\pi}}\int_{-S}^{+S} dt\,e^{-\beta t^2}\leq\|\hat{O}_0\| e^{-\beta S^2}+C_{H_{\partial\Omega} O_0}Se^{\frac{vS-R}{\xi_0}}\notag\\
&\leq&\left(\|\hat{O}_0\|+C_{H_{\partial\Omega} O_0}\tfrac{\sqrt{2}R}{\epsilon\Delta\xi_\epsilon}\right)e^{-\frac{R}{\xi_\epsilon}}.
\end{eqnarray}

Using
\begin{eqnarray}
&&\langle\hat{O}^\dagger\hat{Q}_{\text{high}}[\hat{H},\hat{O}]\rangle\leq\sqrt{\langle\hat{O}^\dagger\hat{Q}_{\text{high}}\hat{O}\rangle\langle[\hat{H}_{\partial\Omega},\hat{O}]^\dagger[\hat{H}_{\partial\Omega},\hat{O}]\rangle}\leq2\|\hat{H}_{\partial\Omega}\| \|\hat{O}_0\|\sqrt{\langle\hat{O}^\dagger\hat{Q}_{\text{high}}\hat{O}\rangle},\\
&&\langle\hat{O}^\dagger\hat{Q}_{\text{low}}[\hat{H},\hat{O}]\rangle\leq\langle\hat{O}^\dagger\hat{Q}_{\text{low}}\hat{O}\rangle(1+\epsilon)\Delta,
\end{eqnarray}
and
\begin{eqnarray}
&&\langle\hat{O}^\dagger\hat{Q}\hat{O}\rangle=\langle\hat{O}^\dagger\hat{Q}\hat{O}\rangle+\langle\hat{O}^\dagger\hat{Q}(\hat{O}-\hat{O})\rangle+\langle(\hat{O}-\hat{O})^\dagger\hat{Q}\hat{O}\rangle,\\
&&\langle\hat{O}^\dagger\hat{Q}\hat{O}\rangle-2\|\hat{O}_0\|\|\hat{O}-\hat{O}\|\leq\langle\hat{O}^\dagger\hat{Q}\hat{O}\rangle\leq\langle\hat{O}^\dagger\hat{Q}\hat{O}\rangle+2\|\hat{O}_0\|\|\hat{O}-\hat{O}\|,
\end{eqnarray}
we have
\begin{eqnarray}
\frac{\langle\hat{O}^\dagger\hat{Q}[\hat{H},\hat{O}]\rangle}{\langle\hat{O}^\dagger\hat{Q}\hat{O}\rangle}&=&\frac{\langle\hat{O}^\dagger\hat{Q}_{\text{low}}[\hat{H},\hat{O}]\rangle+\langle\hat{O}^\dagger\hat{Q}_{\text{high}}[\hat{H},\hat{O}]\rangle}{\langle\hat{O}^\dagger\hat{Q}_{\text{low}}\hat{O}\rangle+\langle\hat{O}^\dagger\hat{Q}_{\text{high}}\hat{O}\rangle}\notag\\
&\leq&\frac{\langle\hat{O}^\dagger\hat{Q}_{\text{low}}\hat{O}\rangle(1+\epsilon)\Delta+2\|\hat{H}_{\partial\Omega}\| \|\hat{O}_0\|\sqrt{\langle\hat{O}^\dagger\hat{Q}_{\text{high}}\hat{O}\rangle}}{\langle\hat{O}^\dagger\hat{Q}_{\text{low}}\hat{O}\rangle}\notag\\
&\leq&(1+\epsilon)\Delta+2\|\hat{H}_{\partial\Omega}\| \|\hat{O}_0\|\frac{\sqrt{\langle\hat{O}^\dagger\hat{Q}_{\text{high}}\hat{O}\rangle+2\|\hat{O}_0\|\|\hat{O}-\hat{O}\|}}{\langle\hat{O}^\dagger\hat{Q}_{\text{low}}\hat{O}\rangle-2\|\hat{O}_0\|\|\hat{O}-\hat{O}\|}\notag\\
&\leq&(1+\epsilon)\Delta+2\|\hat{H}_{\partial\Omega}\|e^{-\frac{R}{2\xi_\epsilon}} \frac{\sqrt{\|\hat{O}_0\|+2\left(\|\hat{O}_0\|+C_{H_{\partial\Omega} O_0}\tfrac{\sqrt{2}R}{\epsilon\xi_\epsilon\Delta}\right)}}{w-2\left(\|\hat{O}_0\|+L_{H_{\partial\Omega} a_0}\tfrac{\sqrt{2}R}{\epsilon\xi_\epsilon\Delta}\right)e^{-\frac{R}{\xi_\epsilon}}}.
\end{eqnarray}
Therefore, $\hat{O}$ has the property stated in the main text.

\end{document}